\documentclass[twocolumn,english,showpacs,preprintnumbers,superscriptaddress,prl]{revtex4}
\usepackage{ae,aecompl}
\usepackage[T1]{fontenc}
\usepackage[latin1]{inputenc}
\usepackage{verbatim}
\usepackage{textcomp}
\usepackage{amsmath,amsfonts}
\usepackage{graphicx}
\usepackage{amssymb}
\usepackage{multirow}
\makeatletter

\usepackage{subfigure}
\usepackage{dcolumn}\usepackage{bm}\@ifundefined{definecolor}
 {\usepackage{color}}{}
\usepackage{latexsym}

\makeatother

\usepackage{babel}
\definecolor{Red}{rgb}{1,0.0,0.0}
\begin{document}

\title{ Robust dynamical decoupling for quantum computing and quantum memory}

\author{Alexandre M. Souza}

\author{Gonzalo A. \'{A}lvarez}

\author{Dieter Suter}

\affiliation{Fakult\"at Physik, Technische Universit\"at Dortmund, D-44221,
Dortmund, Germany}

\date{\today}
\begin{abstract}

Dynamical decoupling (DD) is a popular technique for protecting qubits from the environment.
However, unless special care is taken, experimental errors in the control pulses 
used in this technique can destroy
the quantum information instead of preserving it.
Here, we investigate techniques for making DD sequences robust against
different types of experimental errors while retaining good decoupling efficiency
in a fluctuating environment. We present experimental data from solid-state nuclear spin 
qubits and introduce a new DD sequence
that is suitable for quantum computing and quantum memory.

\end{abstract}

\maketitle

An obstacle against  high-precision quantum
control is the decoherence process \cite{zurek}. The ability
to preserve quantum behavior in the presence 
of noise is essential for the  performance of quantum devices,
such as quantum memories \cite{epjd}, where one wishes to store a quantum 
state, and quantum computers, where the quantum information is processed  \cite{nielsen}.

A promising strategy developed to avoid decoherence is the dynamical
decoupling (DD) method \cite{viola}, which aims to reduce
decoherence times by attenuating
the system-environment interaction. 
Since DD does not require
auxiliary qubits or measurements, it can be used as an economical alternative
to complement quantum error correcting codes \cite{viola3}.
Decoupling schemes were originally developed in the framework of
nuclear magnetic resonance (NMR)  \cite{waugh1}. 
In DD, a sequence
of control fields is periodically
applied to a system in cycles of period $\tau_{c}$, in order to
refocus the system-environment evolution. 
The delay  $\tau_d$ between pulses is one of the relevant parameters 
of a sequence \cite{pra}. When $\tau_{d}$ is shorter than the correlation
time  $\tau_{e}$ of the environment, the preservation of a single
qubit state is ideally possible even for the most general dephasing 
environment \cite{cdd}. The effectiveness 
of a decoupling scheme depends crucially on the 
repetition rate with which the pulses can be applied. 
Recent experiments have successfully implemented
DD methods and demonstrated the resulting increase of the coherence times \cite{pra,exp2,exp3}.
However, systems with fast fluctuating environment, with short $\tau_{e}$,
put enormous demands on the hardware.
They may be encountered in a wide range of quantum information 
processing implementations \cite{revnat} and
represent the most challenging regime for avoiding decoherence.

Apart from preserving the state of a quantum memory as long as possible, 
DD can also be used  to keep a quantum state coherent while logical operations
are performed on it.  In this case, one needs to refocus the system-environment 
interactions while not refocusing the desired system-system interactions needed for 
multi-qubit gates, as discussed in Ref. \cite{sutergate}. Suitable DD sequences for quantum 
computing  should therefore  keep the number of refocusing pulses small to allow the 
computational operations and limit power deposition on the sample.
Since the precision of any real operation is 
finite, the performance of experimentally accessible DD sequences is limited also by the 
pulse errors \cite{pra,exp2,aperr}. This effect must be minimized by choosing a decoupling 
scheme that is robust against pulse imperfections.

As the spacing between the pulses is reduced, the refocusing gets more effective, since the state
of the environment appears more static on the time scale of the pulse spacing.
Ideally, this can be extended indefinitely, until the system decouples completely from the environment
in the limit of infinitely short pulse spacing.
In practice, however, the observed decay rate goes through a minimum and subsequently increases again.
In this regime, the signal is destroyed primarily by imperfections of the refocusing pulses.

We first consider this limiting case, where the effect of the environment is negligible and we need to
discuss only the effect of control errors.
The dominant cause of these errors is, in most cases, a deviation between the actual and the ideal 
amplitude of the control field. The result of this amplitude error is that the rotation 
angle deviates from $\pi$, typically by a few percent. A possible approach for compensating these 
errors is the use of composite pulses \cite{comp},
which generate rotations that are close to the target value even in the presence of amplitude errors.
In this case, the error correction is done ``inside" the pulse.
Alternatively, it is possible to design the sequence in such a way that the error introduced by one pulse is compensated
by subsequent pulses. 
We refer to the former approach as using robust pulses and to the second approach
as using robust or self-correcting sequences.

\begin{figure}[htbp]
\vspace*{13pt}
\begin{center}
{\includegraphics[width=6.0cm]{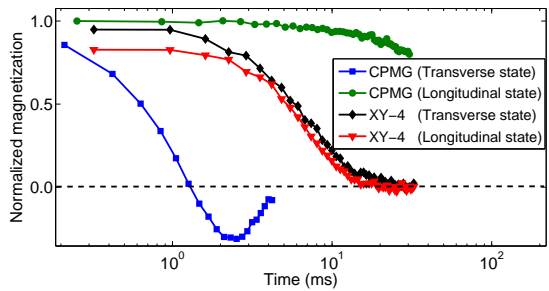}}
\end{center}
\vspace*{13pt}
\caption{\label{figure1a}   Normalized magnetization obtained experimentally for two basic NMR sequences: one non-robust 
against errors, CPMG, and one self-correcting sequence, XY-4.}
\end{figure}

Apart from reducing the coherence time in a certain parameter range, pulse imperfections also make
some DD sequences asymmetric with respect to the initial conditions.
We illustrate this in Fig. \ref{figure1a}, where we plot the signal of the carbon nuclear spin magnetization 
in the Adamantane molecule \cite{pra} measured  for two different DD sequences 
and initial conditions.
Considering first the Carr-Purcell-Meiboom-Gill (CPMG) sequence, which 
consists of identical $\pi$-pulses.
It was originally designed to
preserve a single component of the magnetization \cite{mg}.
Here, the decoherence time is $\approx 100$ ms if the Bloch vector of the qubit is initially oriented 
parallel to the pulse axis  (longitudinal state).
Under these conditions, pulse errors do not affect the coherence \cite{mg} - they even provide additional stabilization \cite{pra,belen,libarret}.
In contrast, if the initial condition is perpendicular to the direction of the 
pulses  (transverse state; blue squares in Fig. \ref{figure1a}), 
the errors of the individual pulses accumulate and lead to a rapid decay: the signal completely 
vanishes after $\approx 50$ pulses ($\approx 5$ ms).
A similar behavior is found for the UDD sequence \cite{udd}, which also uses rotations around a single axis \cite{ashok}.

The second DD sequence represented in Fig. \ref{figure1a} is the XY-4 
sequence \cite{xy}, which consists of  $\pi$ pulses 
applied along the $x$ and $y$ axes 
%(see Table \ref{tab1}). 
It performs much more symmetrically with respect 
to the initial condition: both initial states decay on a time-scale of $\approx 10$ ms. 
The observed time scale also shows that this sequence has a built-in partial error compensation.

\begin{figure}[htbp]
\vspace*{13pt}
\begin{center}
{\includegraphics[width=6.0cm]{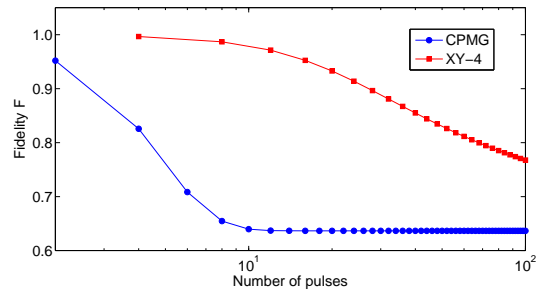}}
\end{center}
\vspace*{13pt}
\caption{\label{figure1b} 
Simulation of fidelity decay due to pulse errors. 
The blue circles represent the decay of the fidelity due  only to pulse 
errors for the 
CPMG sequence and the red squares for the XY-4 sequence. 
The fidelity is calculated as an average over  a Gaussian distribution of flip angles; the width 
was 10\% of the nominal flip angle. }
\end{figure}

For preserving an unknown quantum state, the appropriate performance measure should not
depend on the initial condition. A common choice for quantifying the performance of quantum operations is then 
the fidelity 
\begin{equation}
F = \frac{|Tr(A B^{\dag})|}{\sqrt{Tr(AA^{\dag}) Tr(BB^{\dag})}}.
\label{fidel}
\end{equation}
Here, $A$ is the target propagator (unity for the examples discussed here) and $B$ is the actual propagator
generated by the real pulse sequence.
In Fig. \ref{figure1b}, we numerically simulate the fidelity decay during the two DD sequences discussed above 
due  only to errors
in the flip angles of the pulses, neglecting  environmental effects.
For the CPMG sequence (blue circles), we observe a rapid decay to the limiting fidelity of $\approx 0.65$.
At this point, the system reaches the completely disordered state $\rho \propto \mathbf{1}$,
with the order dephased in the inhomogeneous field distribution.
 The fast decay  for CPMG is 
experimentally manifested by the large asymmetry shown in Figure \ref{figure1a}. The XY-4 sequence, however, 
represented by the red squares in Fig. \ref{figure1b}, causes  a slower decay for the same parameters, 
indicating a partial error compensation over the cycle.

In order to correct errors of non-robust sequences or even to improve the error 
tolerance of self-correcting sequences, it is possible to replace the simple
$\pi$ pulses by
composite pulses \cite{comp} that are more robust against errors. 
We tested different  class-A composite $\pi$-pulses, which produce compensated 
rotations for any initial condition, and found that the pulse used in Ref.
\cite{exp2} 
\begin{equation}
(\pi)_{\pi/6+\phi}-(\pi)_{\phi}-(\pi)_{\pi/2+\phi}-(\pi)_{\phi}-(\pi)_{\pi/6+\phi},\label{kpul}
\end{equation}
to which we refer here as the Knill pulse, is the most robust against flip-angle
errors and off-resonance errors, which are the leading errors in many experimental situations.
The Knill pulse
is equivalent to a robust $\pi$ rotation around the axis defined by $\phi$ followed by a $-\pi/3$
rotation around the $z$ axis. 
For cyclic sequences, which always consist of even numbers of $\pi$ rotations, the effect 
of the additional $z$ rotation vanishes  if the flip angle errors are sufficiently low. 
We inserted these composite pulses into different DD sequences to test their performance.

Since the Knill pulse consists only of $\pi$ rotations, it can also be used as a DD sequence:
instead of concatenating the pulses directly, we inserted delays between them and obtained thus a
new DD sequence
with the rotation axes given by Eq. (\ref{kpul}):
KDD$_\phi$ = $\tau/2-(\pi)_{\pi/6+\phi}-\tau-(\pi)_{\phi}-\tau-(\pi)_{\pi/2+\phi}-\tau-(\pi)_{\phi}-\tau-(\pi)_{\pi/6+\phi}-\tau/2$. 
To further improve 
the robustness of the sequence,  we also extended it by combining 5-pulse 
blocks shifted in phase by $\pi/2$,
such as [KDD$_\phi$ - KDD$_{\phi+\pi/2}$]$^2$, where the lower index
gives the overall phase of the block. 
We will refer to the cyclic repetition of these 20 pulses as the KDD sequence.

\begin{figure}[htbp]
\vspace*{13pt}
\begin{center}
{\includegraphics[width=8.5cm]{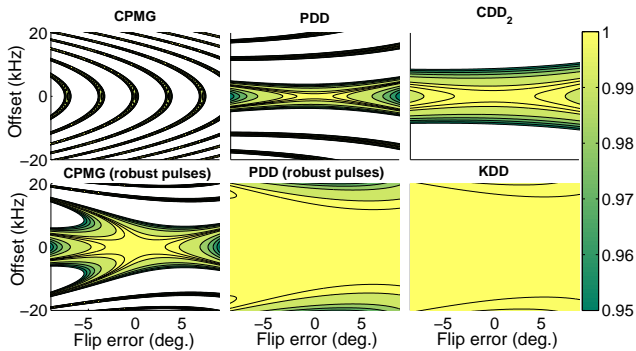}}
\end{center}
\vspace*{13pt}
\caption{\label{figure2}  Error tolerance of different  DD sequences. 
The upper row shows the calculated  fidelity $F$ for standard DD sequences, while the lower row shows the results for
the CPMG and PDD=XY-4 sequences when the $\pi$ pulses are replaced by Knill pulses.
The last panel corresponds to the KDD sequence, also based on the Knill pulse.
Each panel shows the fidelity after 100 pulses as a function of flip angle 
error and offset errors. 
The regions where the 
fidelity is lower than 0.95 are shown in white. 
The highest contour level is $F = 0.999.$ }
\end{figure}

Fig. \ref{figure2} summarizes how sensitive the different DD sequences are to the two main 
pulse imperfections  (no dephasing environment):
Each panel shows the simulated fidelity $F$ after 100 pulses with a combination of frequency (vertical axis) and amplitude errors
(horizontal axis).
Fidelities $>0.95$ are color coded.
For CPMG (first panel), the fidelity after 100 pulses drops to <0.95 even for very small flip angle errors
or offsets. 
The concatenated sequences (CDD-$n$) were constructed as $[CDD_{n-1}-X-CDD_{n-1}-Y]^2$ 
and $[\sqrt{CDD_{n-1}}-X-CDD_{n-1}-Y-\sqrt{CDD_{n-1}}]^2$ for the standard and symmetrized 
versions respectively. For $n=1$, we have
$CDD_1$ = XY-4  = PDD. The basic sequence XY-4 is defined as $[\tau_d-X-\tau_d-Y]^2$ for 
the standard case. The symmetrized sequence is  $[\tau_d/2-X-\tau_d-Y-\tau_d/2]^2$.
The $2^{nd}$ and $3^{rd}$ panel in Fig. \ref{figure2} show the corresponding results for the
 standard (asymmetric) XY-4 and CDD$_2$ sequence. 
 Clearly, they are much less susceptible to flip angle errors than the CPMG sequence.
As shown in the lower panels, a further significant improvement is achieved if the $\pi$-rotations are replaced by Knill pulses,
at the expense of increased power deposition (by a factor of 5).
The best performance, with similar power deposition as for the panels in the upper row,
is achieved with the KDD sequence.

Now we start looking at the effect of imperfect DD pulses on 
a fluctuating environment. We tested the sequences discussed above in an experimental setting.
As the system qubit, we used $^{13}$C nuclear spins in the CH$_2$ groups 
of a polycrystalline Adamantane sample. The natural abundance carbon spins are surrounded by 
$^{1}$H nuclear spins acting as a rapidly fluctuating environment  \cite{pra}.
The bath correlation time was  $\approx 100 \mu s$, the pulse length was $10.6 \mu s$ and the 
delays between the pulses were varied from $5 \mu s$ to $150 \mu s$. 
Under our conditions, the interaction between the carbon
nuclei can be neglected and the decoherence mechanism is a 
pure dephasing process \cite{pra}. 
The experiments were performed on a home-built 300 MHz solid-state NMR spectrometer. 

\begin{figure}[htbp]
\begin{centering}
{\includegraphics*[width=9.0cm]{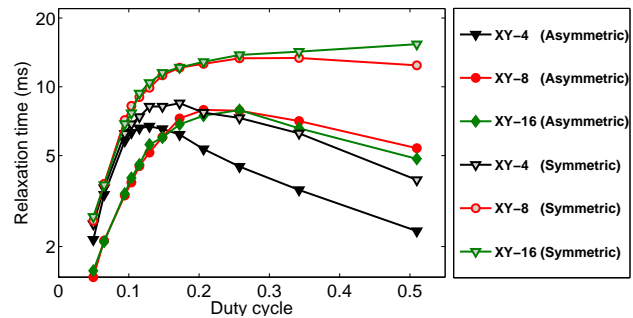}}
\par\end{centering}
\caption{\label{figure3a} 
 Comparison of the decoupling performance of symmetric vs. asymmetric 
XY sequences.}
\end{figure}

\begin{figure}[htbp]
\begin{centering}
{\includegraphics*[width=9.0cm]{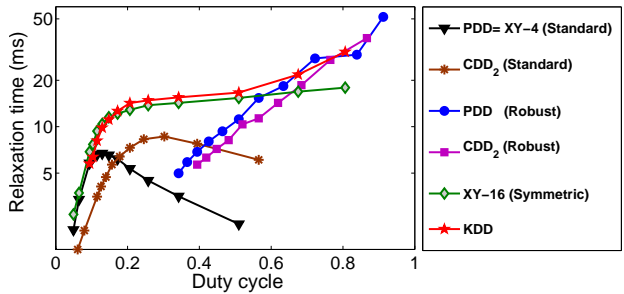}}
\par\end{centering}
\caption{\label{figure3b}  Comparison  of the decoupling performance of standard DD sequences (PDD, CDD$_2$)
 with the same sequences using composite pulses and the self-correcting sequences
XY-16 and KDD.}
\end{figure}

 Figures \ref{figure3a} and \ref{figure3b} summarize the experimental results
by plotting the relaxation times as a function of the duty cycle
(total irradiation time divided by total time). The relaxation times are 
defined as the $1/e$ decay time of the magnetization. In the first set of 
experiments, Fig. \ref{figure3a},  we compare different members of the 
XY family. The sequences were constructed, as explained in \cite{xy}, using 
the standard definition of XY-4  and the its symmetric definition, introduced above.
For all sequences, the symmetric version performs significantly better, with 
relaxation times $\approx 5$ times longer
than for the asymmetric version, irrespective of the initial condition.
This can be attributed to the fact that in the symmetric sequences, all odd-order terms of the 
Magnus expansion vanish \cite{nmr}.
This is a significant advantage, considering that the power deposition and the complexity of 
the sequences are identical.

In the second set of experiments, Fig. \ref{figure3b}, 
we compare standard CDD sequences against CDD sequences with robust pulses
and symmetric timing.
For low duty
cycles, standard CDD sequences perform better.
This is due to the shorter cycle time of the standard sequence
if constant duty cycles are compared.
Using non-robust pulses may therefore be the preferred option if DD sequences
are applied parallel to gate operations.
However
their performance always saturates or decreases with increasing duty cycle under the 
present experimental conditions, while the performance 
of sequences with robust pulses continuous to improve.
Thus, if only the preservation of a quantum state is required, without 
considering power deposition, the sequences with robust
pulses provide the best performance. The performance of the sequences with robust pulses 
shows no significant dependence on the concatenation level. Apparently, the additional 
corrections due to the concatenation are not required in this case.

 For a large range of duty cycles, we find that KDD provides the best performance.
At low duty cycles the performance of KDD is comparable to that of 
self-correcting sequences without robust pulses, indicating that the errors of the 
individual pulses compensate over a cycle. 
For high duty cycles, instead of saturating, the relaxation 
time continues to increase, as in the case of sequences with robust pulses. 
This shows that KDD has suitable properties for computing and memory applications. 

In summary, we have considered the problem of protecting a qubit in
the presence a fast fluctuating spin-bath and pulse imperfections.
Different strategies for correcting
pulse errors were investigated and verified experimentally in DD experiments.  
We observed 
that the symmetrization of sequences 
is an important feature since this leaves the power deposition and the complexity of the sequence constant
but always leads to better performance.
The
best sequences that are suitable for parallel application of quantum gate
operations are the symmetric self-correcting sequences. 
However, their performance saturates at higher duty cycle,
while the performance of sequences with robust pulses continues to improve 
under our experimental conditions.
Thus, if the objective is only to preserve a quantum state, the best performance
is obtained at high duty cycles, using robust pulses.
We also introduced a new DD sequence, which combines the useful properties
of robust sequences with those of robust pulses and can thus be used for both quantum computing and state preservation. 
This new sequence contains rotations around different axes in the $xy$ plane, not only around the more conventional
directions $x$ and $y$. 
We believe that mixing nontrivial  directions  will be a helpful ingredient for
developing new robust dynamical decoupling sequences for future quantum information applications.

\begin{acknowledgments}
We acknowledge useful discussions with Daniel Lidar.
This work is supported by the DFG through Su 192/24-1. G.A.A. thanks
the Alexander von Humboldt Foundation. 
\end{acknowledgments}

%\bibliography{dd}

\begin{thebibliography}{19}
\expandafter\ifx\csname natexlab\endcsname\relax\def\natexlab#1{#1}\fi
\expandafter\ifx\csname bibnamefont\endcsname\relax
  \def\bibnamefont#1{#1}\fi
\expandafter\ifx\csname bibfnamefont\endcsname\relax
  \def\bibfnamefont#1{#1}\fi
\expandafter\ifx\csname citenamefont\endcsname\relax
  \def\citenamefont#1{#1}\fi
\expandafter\ifx\csname url\endcsname\relax
  \def\url#1{\texttt{#1}}\fi
\expandafter\ifx\csname urlprefix\endcsname\relax\def\urlprefix{URL }\fi
\providecommand{\bibinfo}[2]{#2}
\providecommand{\eprint}[2][]{\url{#2}}

\bibitem[{\citenamefont{Zurek}(2003)}]{zurek}
\bibinfo{author}{\bibfnamefont{W.~H.} \bibnamefont{Zurek}},
  \bibinfo{journal}{Rev. Mod. Phys.} \textbf{\bibinfo{volume}{75}},
  \bibinfo{pages}{715} (\bibinfo{year}{2003}).

\bibitem[{\citenamefont{Simon et~al.}(2010)\citenamefont{Simon, et~al.}}]{epjd}
\bibinfo{author}{\bibfnamefont{C.}~\bibnamefont{Simon}},
  \bibnamefont{et~al.}, \bibinfo{journal}{Eur. Phys. J. D}
  \textbf{\bibinfo{volume}{58}}, \bibinfo{pages}{1} (\bibinfo{year}{2010}).

\bibitem[{\citenamefont{Nielsen and Chuang}(2000)}]{nielsen}
\bibinfo{author}{\bibfnamefont{M.~A.} \bibnamefont{Nielsen}} \bibnamefont{and}
  \bibinfo{author}{\bibfnamefont{I.~L.} \bibnamefont{Chuang}},
  \emph{\bibinfo{title}{Quantum Computation and Quantum Information}}
  (\bibinfo{publisher}{Cambridge University Press},
  \bibinfo{address}{Cambridge}, \bibinfo{year}{2000});
  \bibinfo{author}{\bibfnamefont{I.~S.} \bibnamefont{Oliveira}},
    \bibnamefont{et~al.},
\emph{\bibinfo{title}{NMR Quantum information
  processing}} (\bibinfo{publisher}{Elsevier}, \bibinfo{address}{Amsterdam},
  \bibinfo{year}{2007}).

\bibitem[{\citenamefont{Viola et~al.}(1999)\citenamefont{Viola, Knill, and
  Lloyd}}]{viola}
\bibinfo{author}{\bibfnamefont{L.}~\bibnamefont{Viola}},
  \bibinfo{author}{\bibfnamefont{E.}~\bibnamefont{Knill}}, \bibnamefont{and}
  \bibinfo{author}{\bibfnamefont{S.}~\bibnamefont{Lloyd}},
  \bibinfo{journal}{Phys. Rev. Lett.} \textbf{\bibinfo{volume}{82}},
  \bibinfo{pages}{2417} (\bibinfo{year}{1999});
  \bibinfo{author}{\bibfnamefont{W.}~\bibnamefont{Yang}},
  \bibinfo{author}{\bibfnamefont{Z.-Y.} \bibnamefont{Wang}}, \bibnamefont{and}
  \bibinfo{author}{\bibfnamefont{R.-B.} \bibnamefont{Liu}},
  \bibinfo{journal}{Front. Phys.} \textbf{\bibinfo{volume}{6}},
  \bibinfo{pages}{1} (\bibinfo{year}{2010});
  \bibinfo{author}{\bibfnamefont{J.~R.} \bibnamefont{West}},
  \bibinfo{author}{\bibfnamefont{B.~H.} \bibnamefont{Fong}}, \bibnamefont{and}
  \bibinfo{author}{\bibfnamefont{D.~A.} \bibnamefont{Lidar}},
  \bibinfo{journal}{Phys. Rev. Lett.} \textbf{\bibinfo{volume}{104}},
  \bibinfo{pages}{130501} (\bibinfo{year}{2010}).




\bibitem[{\citenamefont{Khodjasteh et~al.}(2010)\citenamefont{Khodjasteh,
  Lidar, and Viola}}]{viola3}
\bibinfo{author}{\bibfnamefont{K.}~\bibnamefont{Khodjasteh}},
  \bibinfo{author}{\bibfnamefont{D.~A.} \bibnamefont{Lidar}}, \bibnamefont{and}
  \bibinfo{author}{\bibfnamefont{L.}~\bibnamefont{Viola}},
  \bibinfo{journal}{Phys. Rev. Lett.} \textbf{\bibinfo{volume}{104}},
  \bibinfo{pages}{090501} (\bibinfo{year}{2010});
  \bibinfo{author}{\bibfnamefont{K.}~\bibnamefont{Khodjasteh}} \bibnamefont{and}
  \bibinfo{author}{\bibfnamefont{L.}~\bibnamefont{Viola}},
  \bibinfo{journal}{Phys. Rev. Lett.} \textbf{\bibinfo{volume}{102}},
  \bibinfo{pages}{080501} (\bibinfo{year}{2009}).


\bibitem[{\citenamefont{Waugh}(1982{\natexlab{a}})}]{waugh1}
\bibinfo{author}{\bibfnamefont{J.~S.} \bibnamefont{Waugh}},
  \bibinfo{journal}{J. Magn. Reson.} \textbf{\bibinfo{volume}{50}},
  \bibinfo{pages}{30} (\bibinfo{year}{1982}{\natexlab{a}});
  \bibinfo{author}{\bibfnamefont{J.~S.} \bibnamefont{Waugh}},
  \bibinfo{journal}{J. Magn. Reson.} \textbf{\bibinfo{volume}{49}},
  \bibinfo{pages}{517} (\bibinfo{year}{1982}{\natexlab{b}}).

\bibitem[{\citenamefont{{\'A}lvarez et~al.}(2010)\citenamefont{{\'A}lvarez,
  Ajoy, Peng, and Suter}}]{pra}
\bibinfo{author}{\bibfnamefont{G.~A.} \bibnamefont{{\'A}lvarez}},
  \bibinfo{author}{\bibfnamefont{A.}~\bibnamefont{Ajoy}},
  \bibinfo{author}{\bibfnamefont{X.}~\bibnamefont{Peng}}, \bibnamefont{and}
  \bibinfo{author}{\bibfnamefont{D.}~\bibnamefont{Suter}},
  \bibinfo{journal}{Phys. Rev. A} \textbf{\bibinfo{volume}{82}},
  \bibinfo{pages}{042306} (\bibinfo{year}{2010}).

\bibitem[{\citenamefont{Khodjasteh and Lidar}(2005)}]{cdd}
\bibinfo{author}{\bibfnamefont{K.}~\bibnamefont{Khodjasteh}} \bibnamefont{and}
  \bibinfo{author}{\bibfnamefont{D.~A.} \bibnamefont{Lidar}},
  \bibinfo{journal}{Phys. Rev. Lett.} \textbf{\bibinfo{volume}{95}},
  \bibinfo{pages}{180501} (\bibinfo{year}{2005});
  \bibinfo{author}{\bibfnamefont{K.}~\bibnamefont{Khodjasteh}} \bibnamefont{and}
  \bibinfo{author}{\bibfnamefont{D.~A.} \bibnamefont{Lidar}},
  \bibinfo{journal}{Phys. Rev. A} \textbf{\bibinfo{volume}{75}},
  \bibinfo{pages}{062310} (\bibinfo{year}{2007}).




\bibitem[{\citenamefont{Biercuk et~al.}(2009)\citenamefont{Biercuk, et~al.}}]{exp3}
\bibinfo{author}{\bibfnamefont{M.~J.} \bibnamefont{Biercuk}},
  \bibnamefont{et~al.},
  \bibinfo{journal}{Nature} \textbf{\bibinfo{volume}{458}},
  \bibinfo{pages}{996} (\bibinfo{year}{2009});
  \bibinfo{author}{\bibfnamefont{J.}~\bibnamefont{Du}},
  \bibnamefont{et~al.},
  \bibinfo{journal}{Nature} \textbf{\bibinfo{volume}{421}},
  \bibinfo{pages}{1265} (\bibinfo{year}{2009});
  \bibinfo{author}{\bibfnamefont{G.}~\bibnamefont{deLange}},
  \bibnamefont{et~al.},
  \bibinfo{journal}{Science} \textbf{\bibinfo{volume}{330}},
  \bibinfo{pages}{60} (\bibinfo{year}{2010}).

\bibitem[{\citenamefont{Ryan et~al.}(2010)\citenamefont{Ryan, Hodges, and
  Cory}}]{exp2}
\bibinfo{author}{\bibfnamefont{C.~A.} \bibnamefont{Ryan}},
  \bibinfo{author}{\bibfnamefont{J.~S.} \bibnamefont{Hodges}},
  \bibnamefont{and} \bibinfo{author}{\bibfnamefont{D.~G.} \bibnamefont{Cory}},
  \bibinfo{journal}{Phys. Rev. Lett.} \textbf{\bibinfo{volume}{105}},
  \bibinfo{pages}{200402} (\bibinfo{year}{2010}).

\bibitem[{\citenamefont{Ladd et~al.}(2010)\citenamefont{Ladd, Jelezko,
  Laflamme, Nakamura, et~al.}}]{revnat}
\bibinfo{author}{\bibfnamefont{T.~D.} \bibnamefont{Ladd}},
  \bibnamefont{et~al.},
  \bibinfo{journal}{Nature} \textbf{\bibinfo{volume}{464}}, \bibinfo{pages}{45}
  (\bibinfo{year}{2010}).


\bibitem[{\citenamefont{West et~al.}(2010)\citenamefont{West, et~al.}}]{sutergate}
\bibinfo{author}{\bibfnamefont{J.~R.} \bibnamefont{West}},
  \bibnamefont{et~al.},
 \bibinfo{note}{arXiv:0911.2398v2}.


\bibitem[{\citenamefont{Wang and Dobrovitski}()}]{aperr}
\bibinfo{author}{\bibfnamefont{Z.-H.} \bibnamefont{Wang}} \bibnamefont{and}
  \bibinfo{author}{\bibfnamefont{V.}~\bibnamefont{Dobrovitski}},
  \bibinfo{note}{arXiv:1101.0292 (Accepted in J. Phys. B)}.


\bibitem[{\citenamefont{Levitt}()}]{comp}
\bibinfo{author}{\bibfnamefont{M.~H.} \bibnamefont{Levitt}},
  \emph{\bibinfo{title}{Composite pulses}}, \bibinfo{note}{in Encyclopedia of
  NMR,  (Wiley, 1996)}.

\bibitem[{\citenamefont{Meiboom and Gill}(1958)}]{mg}
\bibinfo{author}{\bibfnamefont{S.}~\bibnamefont{Meiboom}} \bibnamefont{and}
  \bibinfo{author}{\bibfnamefont{D.}~\bibnamefont{Gill}},
  \bibinfo{journal}{Rev. Sci. Instrum.} \textbf{\bibinfo{volume}{29}},
  \bibinfo{pages}{688} (\bibinfo{year}{1958}).

\bibitem[{\citenamefont{Franzoni and Levstein}(2005)}]{belen}
\bibinfo{author}{\bibfnamefont{M.B.}~\bibnamefont{Franzoni}} \bibnamefont{and}
  \bibinfo{author}{\bibfnamefont{P.R.}~\bibnamefont{Levstein}},
  \bibinfo{journal}{Phys. Rev. B} \textbf{\bibinfo{volume}{72}},
  \bibinfo{pages}{235410} (\bibinfo{year}{2005}).

\bibitem[{\citenamefont{Li et~al.}(2007)}]{libarret}
\bibinfo{author}{\bibfnamefont{D.} \bibnamefont{Li}},
  \bibnamefont{et~al.},
  \bibinfo{journal}{Phys. Rev. Lett.} \textbf{\bibinfo{volume}{98}},
  \bibinfo{pages}{190401} (\bibinfo{year}{2007}).

\bibitem[{\citenamefont{Uhrig}(2007)}]{udd}
\bibinfo{author}{\bibfnamefont{G.~S.} \bibnamefont{Uhrig}},
  \bibinfo{journal}{Phys. Rev. Lett.} \textbf{\bibinfo{volume}{98}},
  \bibinfo{pages}{100504} (\bibinfo{year}{2007}).

\bibitem[{\citenamefont{Ajoy et~al.}(2011)\citenamefont{Ajoy, \'Alvarez, and
  Suter}}]{ashok}
\bibinfo{author}{\bibfnamefont{A.}~\bibnamefont{Ajoy}},
  \bibinfo{author}{\bibfnamefont{G.~A.} \bibnamefont{\'Alvarez}},
  \bibnamefont{and} \bibinfo{author}{\bibfnamefont{D.}~\bibnamefont{Suter}},
  \bibinfo{journal}{Phys. Rev. A} \textbf{\bibinfo{volume}{83}},
  \bibinfo{pages}{032303} (\bibinfo{year}{2011}).

\bibitem[{\citenamefont{Gullion et~al.}(1990)\citenamefont{Gullion, Baker, and
  Conradi}}]{xy}
\bibinfo{author}{\bibfnamefont{T.}~\bibnamefont{Gullion}},
  \bibinfo{author}{\bibfnamefont{D.~B.} \bibnamefont{Baker}}, \bibnamefont{and}
  \bibinfo{author}{\bibfnamefont{M.~S.} \bibnamefont{Conradi}},
  \bibinfo{journal}{J. Magn. Reson.} \textbf{\bibinfo{volume}{89}},
  \bibinfo{pages}{479} (\bibinfo{year}{1990}).

\bibitem[{\citenamefont{Ernst et~al.}(1987)\citenamefont{Ernst, Bodenhausen,
  and Wokaum}}]{nmr}
\bibinfo{author}{\bibfnamefont{R.~R.} \bibnamefont{Ernst}},
  \bibinfo{author}{\bibfnamefont{G.}~\bibnamefont{Bodenhausen}},
  \bibnamefont{and} \bibinfo{author}{\bibfnamefont{A.}~\bibnamefont{Wokaum}},
  \emph{\bibinfo{title}{Principles of NMR in one and two
  dimensions}} (\bibinfo{publisher}{Clarendon Press}, \bibinfo{address}{Oxford}, \bibinfo{year}{1987}).

\end{thebibliography}

\end{document}